\documentclass[twocolumn]{aastex631}

\usepackage{amsmath} 
\shorttitle{Time lags of inhomogeneous AGN disks}
\shortauthors{Guowei Ren et al.}

\begin{document}

\title{Sizes of active galactic nuclei inhomogeneous disks - large in microlensing, small in reverberation mapping}

\correspondingauthor{Mouyuan Sun}
\email{msun88@xmu.edu.cn}

\author[0000-0002-1497-8371]{Guowei Ren}
\affiliation{Department of Astronomy, Xiamen University, Xiamen, 
Fujian 361005, China; msun88@xmu.edu.cn}

\author[0000-0002-0771-2153]{Mouyuan Sun}
\affiliation{Department of Astronomy, Xiamen University, Xiamen, 
Fujian 361005, China; msun88@xmu.edu.cn}

\author[0000-0002-4419-6434]{Jun-Xian Wang}
\affiliation{CAS Key Laboratory for Research in Galaxies and Cosmology, Department of Astronomy, University of Science and Technology of China, Hefei 230026, China}
\affiliation{School of Astronomy and Space Science, University of Science and Technology of China, Hefei 230026, China}

\author[0000-0002-4223-2198]{Zhen-Yi Cai}
\affiliation{CAS Key Laboratory for Research in Galaxies and Cosmology, Department of Astronomy, University of Science and Technology of China, Hefei 230026, China}
\affiliation{School of Astronomy and Space Science, University of Science and Technology of China, Hefei 230026, China}

\begin{abstract}

Magnetohydrodynamics (MHD) turbulence can drive significant temperature fluctuations in the accretion disk of an active galactic nucleus (AGN). As a result, the disk can be highly inhomogeneous and has a half-light radius larger than the static Shakura \& Sunyaev Disk (SSD), in agreement with quasar microlensing observations. Meanwhile, the accretion-disk sizes can also be determined using continuum reverberation mappings which measure interband cross correlations and time lags. The interband time lags are often understood in the X-ray reprocessing scenario. Here we show that the interband continuum time lags of the X-ray reprocessing of an inhomogeneous disk are similar to or even smaller than those of a static SSD. Consequently, the X-ray reprocessing of an inhomogeneous disk cannot account for the recent continuum reverberation mappings of some Seyfert 1 AGNs, whose measured time lags are larger than those of a static SSD. In contrast to the tight correlation between UV/optical variations, the cross correlation between X-ray and disk emission is rather weak in this model; this behavior is consistent with recent continuum reverberation mappings. Moreover, the time lags in this model are anti-correlated with the amplitude of disk temperature fluctuations. Our results suggest that the temperature fluctuations should be properly considered when modeling interband continuum time lags.

\end{abstract}


\section{introduction} \label{sec:intro}

Active galactic nuclei (AGNs) emit strong non-stellar emission in the Universe. The strong emission is powered by the supermassive black hole (SMBH) accretion. AGNs (including luminous quasars) are found to be variable in a wide range of the electromagnetic spectrum from radio to X-ray and $\gamma$-ray bands \citep[for a review, see, e.g.,][]{Ulrich1997}; the UV–optical variability amplitudes are $\sim 10\%$\textendash$20\%$ \citep[e.g.,][]{Hook1994, MacLeod2010, Sun2018, Suberlak2021, Stone2022}. The UV$/$optical emission of quasars is believed to be produced by the standard thin accretion disk \citep[hereafter SSD;][]{Shakura1973, Novikov1973}. Validating the SSD model is crucial for enhancing our understanding of the gas dynamics surrounding SMBHs and their mass-growth rates. 

Despite the widespread use of the SSD model to explain various AGN observations, the SSD model breaks down in several ways. Some of the issues in which the SSD model fails to explain AGN observations entirely are as follows: (1) the observed half-light radii from quasar microlensing observations are approximately three times larger than the SSD predictions \citep[e.g., ][]{Morgan2010}; (2) the continuum interband time lags of Seyfert 1 AGNs are significantly longer than expectations of the X-ray reprocessing of an SSD (i.e., the X-ray light-travel-time delays) by a factor of $\sim 3$ \citep[e.g.,][]{Fausnaugh2016, Jiang2017}; (3) the color variations in quasars are timescale-dependent \citep[e.g.,][]{Sun2014}; (4) the mean quasar UV spectral energy distributions (SEDs) at $\log (L_{2200}/[\mathrm{erg\ s^{-1}}]) > 45$ (where $L_{2200}$ refers to the quasar luminosity at rest-frame wavelength $2200\ \mathrm{\AA}$) do not depend upon luminosities \citep{Cai2023}, contrary to the SSD model predictions. 

The continuum interband time lags of some quasars are found to be consistent with the SSD model \cite[e.g.,][but see \citealt{Jiang2017}]{Kokubo2018, Homayouni2019, Yu2020}, which is in contrast with the time lags of less-luminous Seyfert 1 AGNs. This issue \citep[as pointed out by][]{Li2021} is hard to be explained by the X-ray reprocessing of a static SSD with or without winds \citep{Sun2019} and non-blackbody emission \citep[][]{Hall2018}. In some cases, the cross-correlations between X-ray and UV/optical variations are weak \citep[e.g.,][]{Edelson2019}, albeit the UV/optical correlations are strong. This result is also apparently inconsistent with the X-ray reprocessing scenario \citep[but see][for a possible explanation]{Panagiotou2022}. Interestingly, the microlensing half-light radius measurements \citep[which are found to be larger than the SSD model; see, e.g.,][]{Morgan2018} seem to be incompatible with the continuum time lags of quasars. 

\cite{Dexter2011} proposed an inhomogeneous disk model to reproduce the observed half-light radius measurements in quasar microlensing observations (i.e., issue \#1) without invoking the X-ray illumination process. In this model, they argue that, at each radius, the effective temperature ($T_{\mathrm{eff}}(t)$) fluctuates around the corresponding effective temperature of the static SSD ($T_{*}$) due to the unavoidable MHD turbulence \citep[e.g.,][]{Balbus1991, Balbus2003}. Moreover, they assume that the fluctuations of $T_{\mathrm{eff}}(t)$ can be described by the damped random walk (DRW) process, and the damping timescale ($\tau$) is radius independent. This model can be generalized to account for other temperature fluctuation processes. For instance, \cite{Cai2016} revised the model of \cite{Dexter2011} by assuming that $\tau$ is a function of the accretion-disk radius and successfully reproduced the observed timescale-dependent color variations in quasars (i.e., issue \#3). The physical meaning of the radius-dependent $\tau$ is straightforward: temperature fluctuations in accretion disks should operate on thermal timescales because of the thermal-energy conservation law of the SSD \citep[see, e.g.,][]{Sun2020}. Both \cite{Dexter2011} and \cite{Cai2016} assume that temperature fluctuations at different radii are independent from each other. Hence, the two models predicate zero interband time lags, which are inconsistent with continuum reverberation mapping observations \citep[e.g.,][]{Fausnaugh2016, Jiang2017}. 

An upgraded inhomogeneous accretion disk model was proposed by \cite{Cai2018}. In this model, they propose that the interband cross-correlations are driven by a common large-scale variation, which mimicks a net result of the propagation and mixing of small-scale fluctuations, rather than by X-ray illumination. Rather than the light travel difference responsible for the interband time lags in the X-ray reprocessing model, the new model of \cite{Cai2018} offers a new origin for the interband time lags as a result of the differential regression of the common fluctuation. With increasing radius, larger damping timescale causes slower regression of the common fluctuation, thus giving rise to the interband time lags. The model of \cite{Cai2018} can reproduce the tight interband correlations and time lags across UV/optical bands and the timescale-dependent color variation of NGC 5548 without invoking the X-ray reprocessing. 

The X-ray reprocessing of the inhomogeneous accretion disk remains unexplored. Are the continuum time lags of this model larger than or consistent with the static SSD? Can this model explain the time-lag observations in Seyfert 1 AGNs or luminous quasars? In this work, we present the continuum time lags of the X-ray reprocessing of the inhomogeneous disk model of \cite{Dexter2011}. The cross correlation between the X-ray and disk emission in this model can be weak, consistent with recent observations. However, while the half-light radius of this model is larger than the static SSD, the corresponding interband time lags are roughly consistent with or even smaller than the SSD predictions. Hence, this model cannot account for the observed time lags in Seyfert 1 AGNs. Moreover, the interband time lags in this model are anti-correlated with the temperature-fluctuation amplitudes. We stress that temperature fluctuations caused by MHD turbulence are expected to exist widely in AGN disks \citep[e.g.,][]{Secunda2023} and should be properly considered in future theoretical disk-size modelings. 

This paper is formatted as follows. In Section~\ref{sec:method}, we describe our calculation of the X-ray reprocessing of the inhomogeneous disk model. In  Section~\ref{sec:results}, we present the disk half-light radii and interband time lags. In Section~\ref{sec:discussion}, we discuss our results. Conclusions are summarized in Section~\ref{sec:conclusion}.

\section{methods} \label{sec:method}

\subsection{Inhomogeneous disk model} \label{sec:Inhomogeneous disk model}

We add the X-ray illumination to the inhomogeneous disk model of \cite{Dexter2011} and a significant fraction of the X-ray emission is thermalized and reprocessed \citep[whose relevant timescales are negligible; e.g.,][]{Nayakshin2002} as disk emission \citep[e.g.,][]{Cackett2007}. In this inhomogeneous disk model, the accretion disk can be divided into many independent zones. The effective temperature fluctuations in each zone can be described by the damped random walk (DRW) process.\footnote{The half-light radii and interband time lags are insensitive to the choice of the temperature fluctuation process.} That is, for each independently fluctuating zone in the accretion disk, the effective temperature ($T_{\mathrm{d}}$) returns to a mean value ($T_{*} $) on a DRW damping timescale ($\tau_{\mathrm{DRW}}$) with a variability amplitude. The mean value of temperature ($T_{*} $) is the effective temperature of an SSD \citep{Shakura1973}, i.e., 
\begin{equation}
\label{eq:1}
T_{*}(r) =\left ( \frac{3GM\dot{M} }{8\pi r^{3}\sigma  }  \right ) ^{1/4}  \\,
\end{equation}
where $G$, $M$, $\dot{M}$, $r$, and $\sigma$ are the gravitational constant, the black-hole mass, the accretion rate, the radius of the accretion disk, ranging from the innermost radius $r_{\mathrm{in}}$ to the outermost radius $r_{\mathrm {out}}$, and the Stefan–Boltzmann constant, respectively. The accretion rate $\dot{M} = \dot{m}\dot{M}_{\mathrm {crit}}$, where $\dot{m}$ is the dimensionless accretion rate; the ciritical accretion rate $\dot{M}_{\mathrm{crit}} = L_{\mathrm{E}}/(\eta c^2)$, where $\eta=0.1$ is the radiative efficiency, $c$ is the speed of light, and $L_{\mathrm{E}}=1.25\times 10^{38}\,(M/M_{\odot})\ \mathrm{erg\ s^{-1}}$ is the Eddington luminosity of an AGN, where $M_{\odot}$ represents the solar mass.

The effective temperature in the inhomogeneous disk is not a monotonic function of radius $r$ but also depends upon azimuth ($\varphi$) and time, i.e.,  
\begin{equation}
\label{eq:2}
T_{\mathrm{d}}\left (r,\varphi,t \right )  =T_{*} \delta f_{\mathrm{d}}\left ( r,\varphi,t \right ),
\end{equation}
where $\delta f_{\mathrm{d}}\left ( r,\varphi,t \right )$ is the normalized temperature fluctuation in each independent zone. It should be noted that each independently fluctuating zone's average luminosity of the inhomogeneous disk is equal to that of the SSD model on the long timescale, i.e., the average value of $\log \delta f_{\mathrm{d}}(r,\varphi, t)$ should merit the following requirement,
\begin{equation}
   \frac{\sigma  \int_{t_{0} }^{t_{\mathrm{n}}} T^{4}_{\mathrm{d}}(r,\varphi, t)dt}{t_{\mathrm{n}}-t_0}   = \sigma T^{4}_{*}(r) \\.
\end{equation}

The logarithmic fluctuation $\log \delta f_{\mathrm{d}}\left ( r,\varphi,t \right )$ of each independent zone is modeled as a DRW process \citep{Zu2013} 
\begin{equation}
    S(\Delta t)_{\log f} = f_{\mathrm{d}}^{2}\exp \left ( {-|\Delta t|/\tau_{\mathrm{DRW}}} \right ),
\end{equation}
where $S$, $\tau_{\mathrm{DRW}}$, $f_{\mathrm{d}}$, and $\Delta t$ represent the kernel function, the damping timescale, the amplitude, and the time interval between two observations, respectively. \cite{Dexter2011} assume that the damping timescale is $\sim 200$ days, corresponding to the observed damping timescales in quasar optical light curves \citep{MacLeod2010}. There is increasing evidence that the $200$-day damping timescale is strongly underestimated \citep[e.g.,][]{Kozowski2017, Suberlak2021, Zhou2024}. 

\cite{Cai2016} proposed that the timescale-dependent color variations in quasars can be reproduced if the damping time scale anti-correlates with radius. In this work, we assume the damping timescale resembles the thermal timescales, i.e., 
\begin{equation} \label{eq:5}
\tau_{\mathrm{DRW}}(r) =\frac{1}{\alpha \Omega _{\mathrm{K}}}, 
\end{equation}
where $\Omega _{\mathrm{K}} = \sqrt{GM/r^{3}  }$ is the Keplerian angular speed and $\alpha = 0.1$ represents the dimensionless viscosity parameter. 

Motivated by the lamppost X-ray reprocessing model, we assume that the inhomogeneous disk is illuminated by the external variable X-ray emission from a point-like corona, which is located above the black hole on its rotational axis with a given scale height $h$ \citep[e.g.,][]{Cackett2007}. In this way, the differences in the light travel time from the corona to various disk regions are responsible for the observed interband time lags. 

In our calculation, the effective temperature variation at each disk zone depends upon the above DRW fluctuations (i.e., Eq.~\ref{eq:2}) and the heating due to the external X-ray illumination. The effective temperature of each independent zone is the result of the joint contribution of the disk and the corona, which is determined by \citep{Cackett2007}
\begin{equation}
\label{eq:6}
    \begin{split}
        \sigma T^4(r, \varphi, t) &= \sigma T^4_{\mathrm{d}}(r, \varphi, t) \\ &+ \frac{(1-A)hL_{\mathrm{x}}(t-\Delta t_{\mathrm{travel}})}{4\pi r^3}, 
    \end{split}
\end{equation}
where $A$, $L_{\mathrm{x}}$, $\Delta t_{\mathrm{travel}}$, and $h$ are the disk albedo, the X-ray luminosity, the light travel time from the point-like corona to the disk, and the corona scale height, respectively. The first term on the right-hand side represents the viscous heating rate per surface area of the disk, and the second term is the external heating rate per surface area due to the X-ray illumination. The light travel time for the face-on case is 
\begin{equation}
\Delta t_{\mathrm{travel}} = \sqrt{(h^2+r^2)} / c.
\end{equation}
The X-ray luminosity $L_{\mathrm{x}}$ is highly variable and can be expressed as $L_{\mathrm{x}}=L_{\mathrm{x},0}\delta f_{\mathrm{c}}$, where $L_{\mathrm{x},0}$ is the average X-ray luminosity and $\delta f_{\mathrm{c}}(t)$ represents the normalized X-ray variability. We can assume that $\log \delta f_{\mathrm{c}}(t)$ also follows the DRW process with a damping timescale of $20$ days and the variability amplitude $f_{\mathrm{c}}$. Hence, the time-dependent effective temperature of an X-ray illuminated inhomogeneous disk can be calculated from Eq.~\ref{eq:2}, Eq.~\ref{eq:6}, and $\delta f_{\mathrm{c}}(t)$. 

Following \citet{Fausnaugh2016}, the ratio between the irradiation of the corona and the disk internal heating at each radius on the long timescale is assumed to be $1/3$, i.e.,
\begin{equation}
\frac{(1-A)h L_{\mathrm{x},0}}{4\pi r^3} =\frac{1}{3}\sigma  T^{4}_{*}(r).
\end{equation}

We assume that the emission of the disk is a perfect blackbody radiation. Hence, the monochromatic luminosity per unit area of each independently fluctuating zone should be $2 B_{\nu}\left (T(r,\varphi,t) \right )$,
where $B_{\nu}$ is the Planck function. The monochromatic luminosity of each layer is 
\begin{equation}
 d L_{\nu,\mathrm{D}}\left ( r,t \right ) =\int_{0}^{2\pi } 2 B_{\nu}\left (T(r,\varphi,t) \right ) rdr d\varphi.
\end{equation}
The monochromatic luminosity of the entire disk is obtained by integrating
\begin{equation}
 L_{\nu,\mathrm{D}}\left ( t \right ) =\int_{r_{\mathrm{in}}}^{r_{\mathrm{out}}} 
 d L_{\nu,\mathrm{D}}\left ( r,t \right ) \\.
\end{equation}
$L_{\nu,\mathrm{D}}\left ( t \right )$ is a function of time ($t$), i.e., 
the light curve of the X-ray illuminated inhomogeneous disk at a given frequency $\nu$.

\subsubsection{Half-light radius}

Quasar microlensing observations essentially measure the half-
light radius ($r_{\mathrm{h}}$, e.g., \citealt{Morgan2010}). The accumulated luminosity of the accretion disk ($L_{\nu,D }(r_{\mathrm{X}}, t)$) can be defined as
\begin{equation}
\label{eq:11}
 L_{\nu,\mathrm{D}}(r_{\mathrm{X}},t) =\int_{r_{\mathrm{in}}}^{r_{\mathrm{X}}} 
 d L_{\nu,\mathrm{D}}\left ( r,t \right )\\.
\end{equation}
Then, the half-light radius ($r_{\mathrm{h}}$) is defined in such a way that $L_{\nu,\mathrm{D}}(r_{\mathrm{h}},t) = L_{\nu,\mathrm{D} }\left ( t \right ) / 2$.

For an SSD, according to Eq. \ref{eq:1} and $kT_*(r_{\lambda}) = hc/ \lambda$, a characteristic radius ($r_{\lambda}$) corresponding to the wavelength $\lambda$ can be calculated for a given monochromatic luminosity $L_{\nu,\mathrm{D}}$ \citep[e.g.,][]{Morgan2010}. The half-light radius of the SSD model can be obtained by $r_{\mathrm{h, SSD}} = 2.44r_{\lambda}$ \citep[e.g.,][]{Morgan2010, Fian2023}. The half-light radius of an inhomogeneous disk should be calculated numerically. Note that we have ensured that the monochromatic luminosity of the inhomogeneous disk model is the same as the SSD model when comparing the half-light radii of the two models.

\subsubsection{Time lag} \label{sec:time lag}
In this work, the time lags between different wavelengths of the X-ray illuminated inhomogeneous disk model are obtained by the interpolated cross-correlation function method (ICCF; \citealt{Gaskell1986, Peterson1998, Peterson2004}) via the public code PyCCF \citep{Sun2018}. The definition of the correlation coefficient function (CCF) is as follows
\begin{equation}
CCF(\delta t) = \frac{E\left \{ \left [ a\left ( t \right )-\left \langle a \right \rangle     \right ]
\left [ b\left ( t+\delta t \right ) -\left \langle b\right \rangle   \right ]   \right \} }{\sigma_{a}\sigma_{b}} ,
\end{equation}
where $a\left ( t \right )$ and $b\left ( t \right )$ are the two time series, and $\sigma_{a}$ and $\sigma_{b}$ represent the standard deviations of two time series. $CCF(\delta t) = 1, -1, 0$ represents that the two time series are perfectly positively correlated, negatively correlated, and have no correlation. The time lag of two light curves is obtained from the ICCF centroid, and the ICCF centroid is calculated by considering the CCF with correlation coefficients $>0.95 \ r_{\mathrm{peak}}$, where $r_{\mathrm{peak}}$ is the peak value of the ICCF. We use the factor $0.95$ rather than the popular one $0.8$ because our simulated data do not have measurement errors (which dilute the cross correlation). We also computed the time lag of the ICCF centroid for $0.8r_{\mathrm{peak}}$, which is roughly consistent with the result for the time lag of $0.95r_{\mathrm{peak}}$. 

For the SSD model, we can calculate the interband time lags. \cite{Tie2018} points out that the CCF method measures the variability-weighted average radius of an accretion disk, i.e., 
\begin{equation}
\label{eq:13}
r_{\mathrm{v},\lambda} = X_{\mathrm{var}} r_{\lambda},
\end{equation}
where $X_{\mathrm{var}} = 5.04$ for the X-ray reprocessing of an SSD. The time lag between the emission of different bands in the SSD model can be obtained by
\begin{equation}
\label{eq:14}
t_{\mathrm{lag}} = \left (r_{\mathrm{v},\lambda1}- r_{\mathrm{v},\lambda2} \right )/c,
\end{equation}
where $r_{\mathrm{v},\lambda1}$ and $r_{\mathrm{v},\lambda2}$ represent the average radii (Eq.~\ref{eq:13}) of two different wavelengths.

\section{results} \label{sec:results}
\begin{figure*}[ht!]
\includegraphics[width=1 \textwidth]{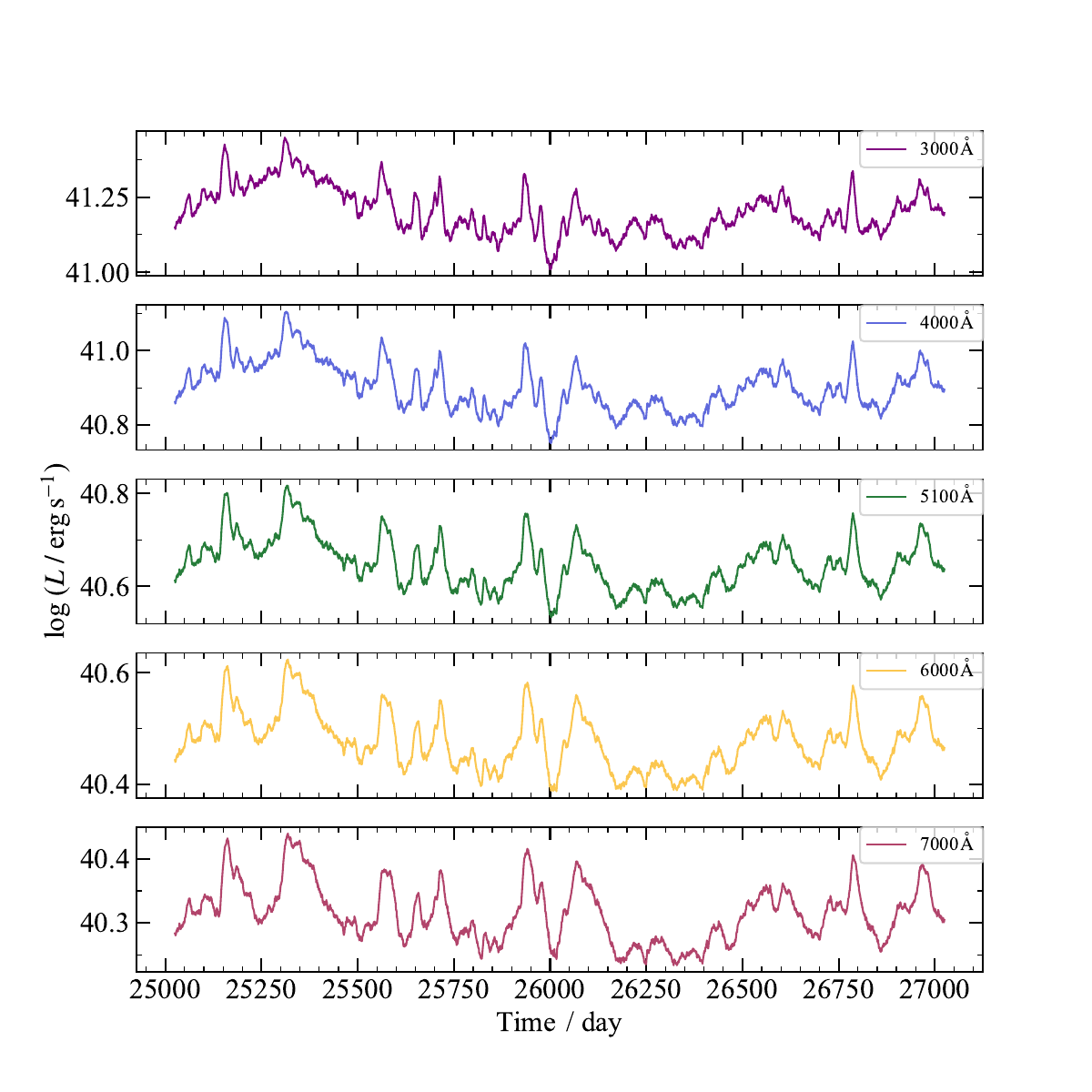}
\caption{The multi-band light curves for the Case 1-1 in Table \ref{tab:Parameters}. }
\label{fig1}
\end{figure*}

For the SSD model and the inhomogeneous disk model, we consider sources with the black-hole mass $\mathrm{log} (M_{\mathrm{BH}}/M_{\odot }) = 7.0, 8.0, 9.0$, the dimensionless accretion rate $\dot{m} = 0.01, 0.1, 0.3$. The accretion disk is divided into numerous zones in $\log{r}$ and $\varphi$ with $N_{\mathrm{r}} = 128$ layers and $N_{\mathrm{\varphi}} = 64$ zones per layer, from the innermost radii $r_{\mathrm{in}} = 3r_{\mathrm{s}}$ to the outermost radii $r_{\mathrm{out}} = 3000r_{\mathrm{s}}$, where $r_{\mathrm{s}} \equiv 2GM /c^{2} $ is the Schwarzschild radius. The cadence and total duration of every simulated light curve are $0.5$ days and $5\times 10^4$ days, respectively. The latter half of the light curves (i.e., 25, 000 days) are used in the subsequent analyses. An example of the multi-band light curves (2000-day long per light curve) for Case 1-1 in Table \ref{tab:Parameters} is shown in Figure \ref{fig1}. To demonstrate our results, we arbitrary fix $f_{\mathrm{d}}=0.3$ and $f_{\mathrm{c}}=0.15$. The parameter settings are shown in Table \ref{tab:Parameters}. We obtain the light curves at rest-frame $3000\ \mathrm{\AA}$, $4000\ \mathrm{\AA}$, $5100\ \mathrm{\AA}$, $6000\ \mathrm{\AA}$, and $7000\ \mathrm{\AA}$. 

\begin{table*}
\caption{Parameters of the SSD model and the inhomogeneous disk model.} 
\label{tab:Parameters} 
\setlength{\tabcolsep}{20mm}{
\begin{tabular}{ccc} 
\hline\hline 
Case ID & $\dot{m}$ &$\mathrm{log} (M_{\mathrm{BH}}/M_{\odot })$  \\
\hline
1-1 & $0.01$   &7.0  \\

1-2 & $0.01$   &8.0  \\

1-3 & $0.01$   &9.0  \\

2-1 & $0.1$    &7.0 \\
2-2 & $0.1$    &8.0  \\
2-3 & $0.1$    &9.0 \\
3-1 & $0.3$    &7.0 \\
3-2 & $0.3$    &8.0  \\
3-3 & $0.3$    &9.0  \\

\hline 
\end{tabular}}
\\
Notes: $\dot{m}$ is the accretion rate; $\mathrm{log} (M_{\mathrm{BH}}/M_{\odot })$ is the logarithmic black hole mass.
\end{table*}

\begin{figure*}[ht!]
\includegraphics[width=1 \textwidth]{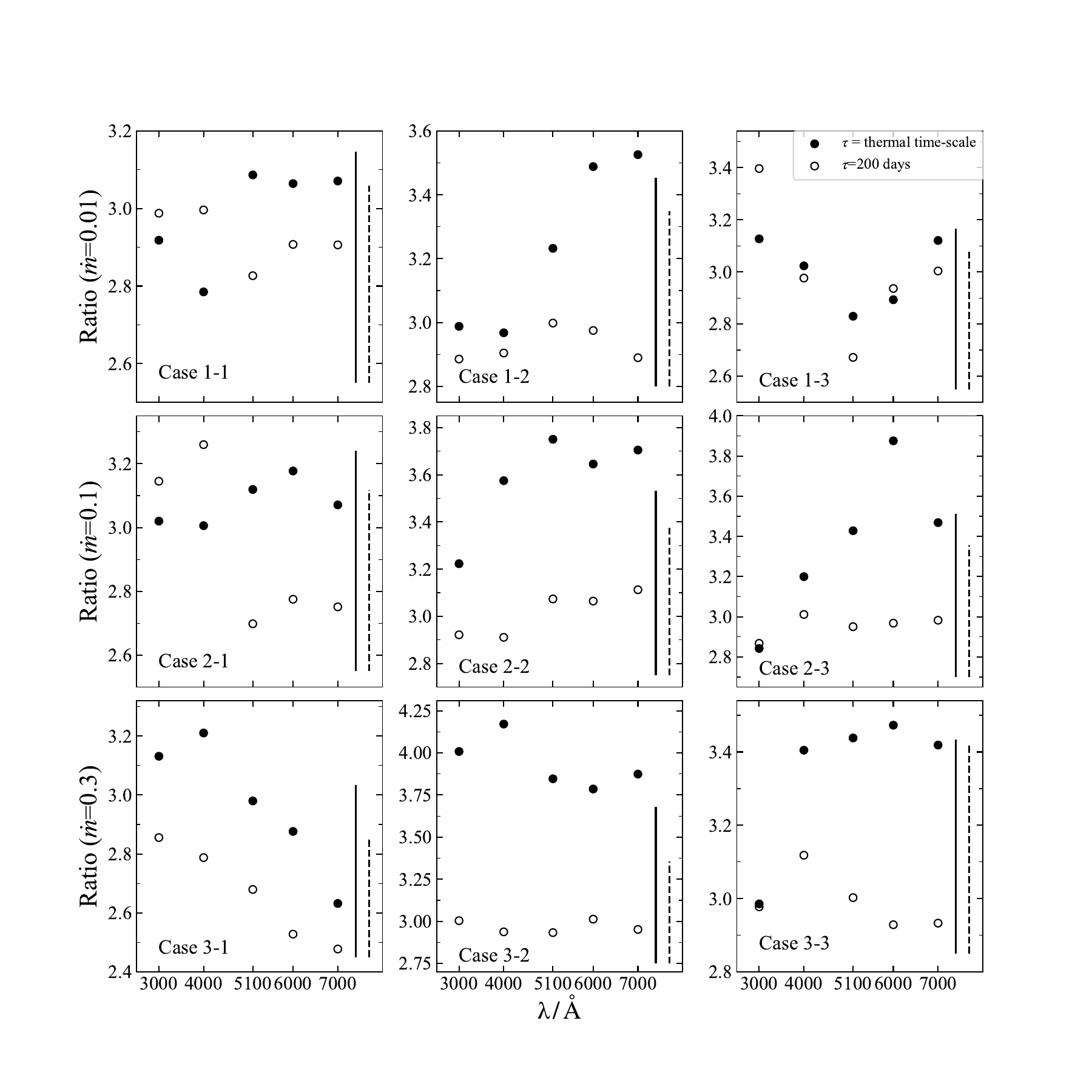}
\caption{The magnification of the accretion disk sizes as a function of the rest-frame wavelength for the nine cases in Table \ref{tab:Parameters}. The black dots correspond to $\tau_{\mathrm{DRW}}=1/(\alpha \Omega_{\mathrm{K}})$ (i.e., the thermal timescale), and the black open circles are for $\tau_{\mathrm{DRW}}= 200$ days. The solid and dashed lines on the right side of each panel are the mean uncertainties for the black dots and the black open circles, respectively. The y-axes represent the ratio of the half-light radius of the inhomogeneous disk model ($r_{\mathrm{h,r}}$) to that of the SSD model ($r_{\mathrm{h, S}}$). For each wavelength, we ensure that the monochromatic luminosity of the inhomogeneous disk model is the same as the SSD model. \label{fig2}} 
\end{figure*}

\begin{figure*}[ht!] 
\includegraphics[width=1 \textwidth]{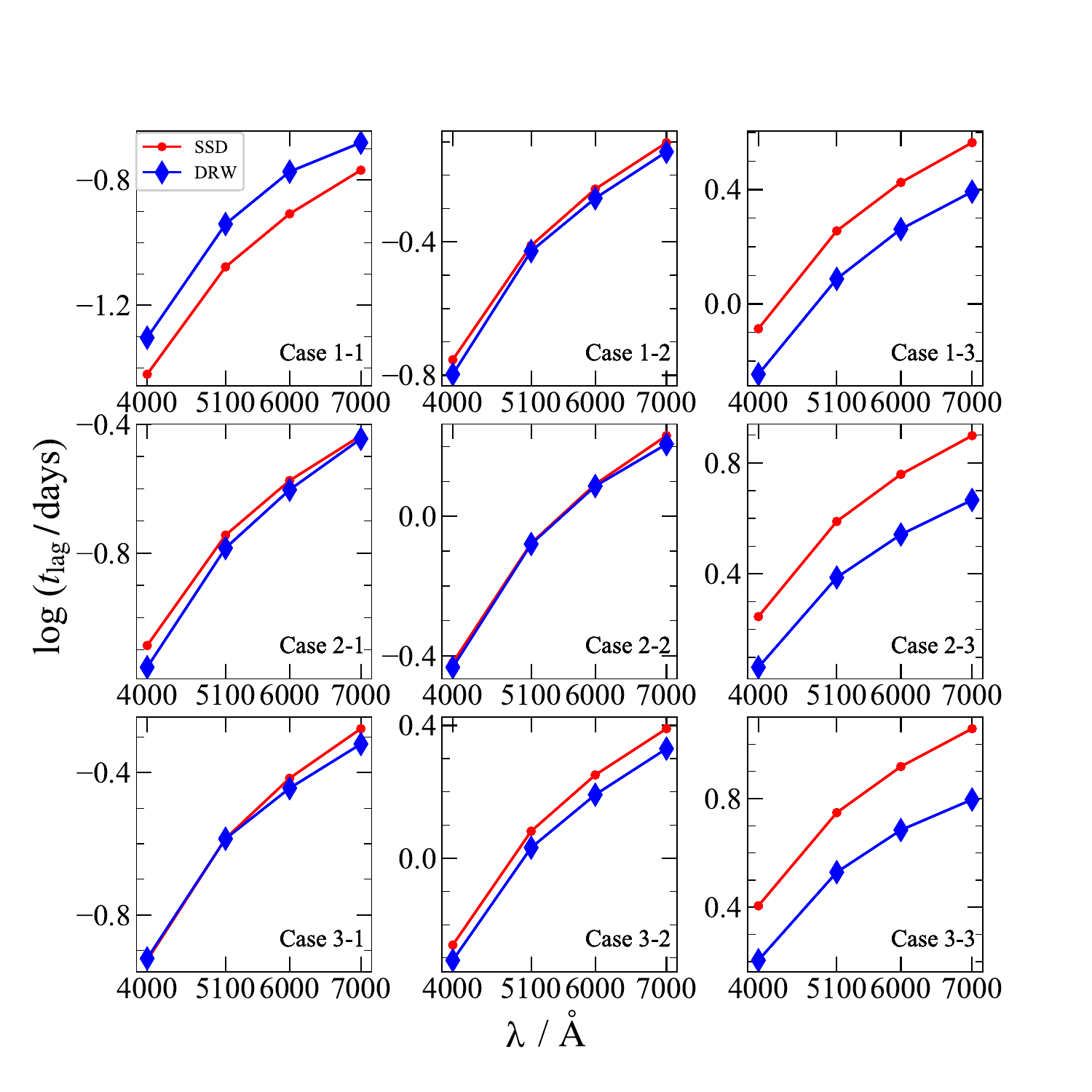}
\caption{Time lags between the 3000 $\textrm{\AA}$ light curve and the light curves in 4000 $\textrm{\AA}$, 5100 $\textrm{\AA}$, 6000 $\textrm{\AA}$ and 7000 $\textrm{\AA}$ for the nine cases in Table \ref{tab:Parameters}. The solid blue and red curves represent the time lags in the X-ray illuminated inhomogeneous and SSD disks. The time lags of the inhomogeneous disk are roughly consistent with or even smaller than those of an SSD disk.\label{fig3}}
\end{figure*}

\begin{figure*}[ht!]
\includegraphics[width=1 \textwidth]{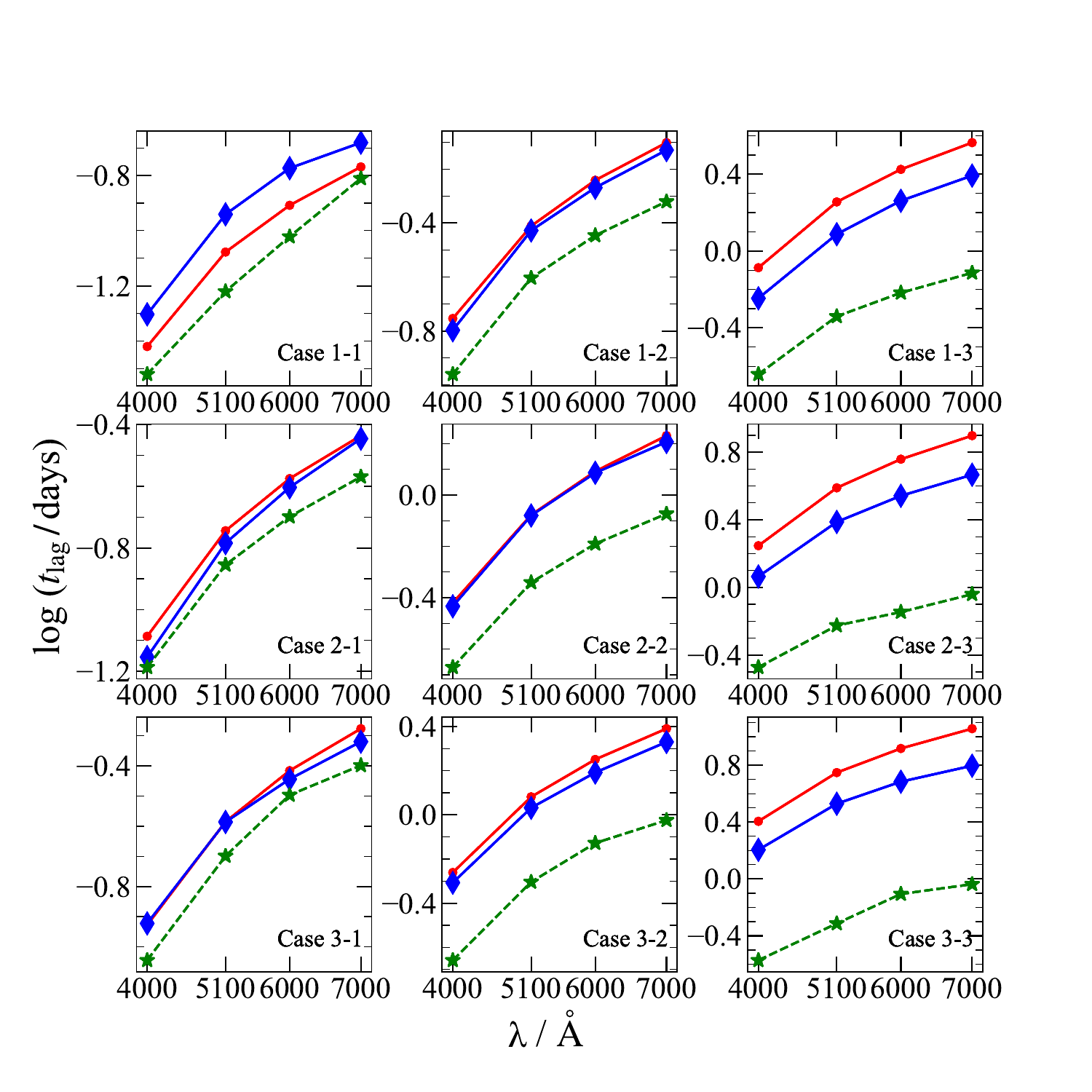}
\caption{Time lags as a function of wavelengths. The blue and red curves and symbols are taken from Figure \ref{fig3}. The green stars represent the inhomogeneous disk time lags for the case of $\tau =200$ days. For comparison, the results for the blue curves correspond to $\tau=1/(\alpha \Omega_{\mathrm{K}})$. \label{fig4}}
\end{figure*}

\begin{figure*}[ht!]
\includegraphics[width=1 \textwidth]{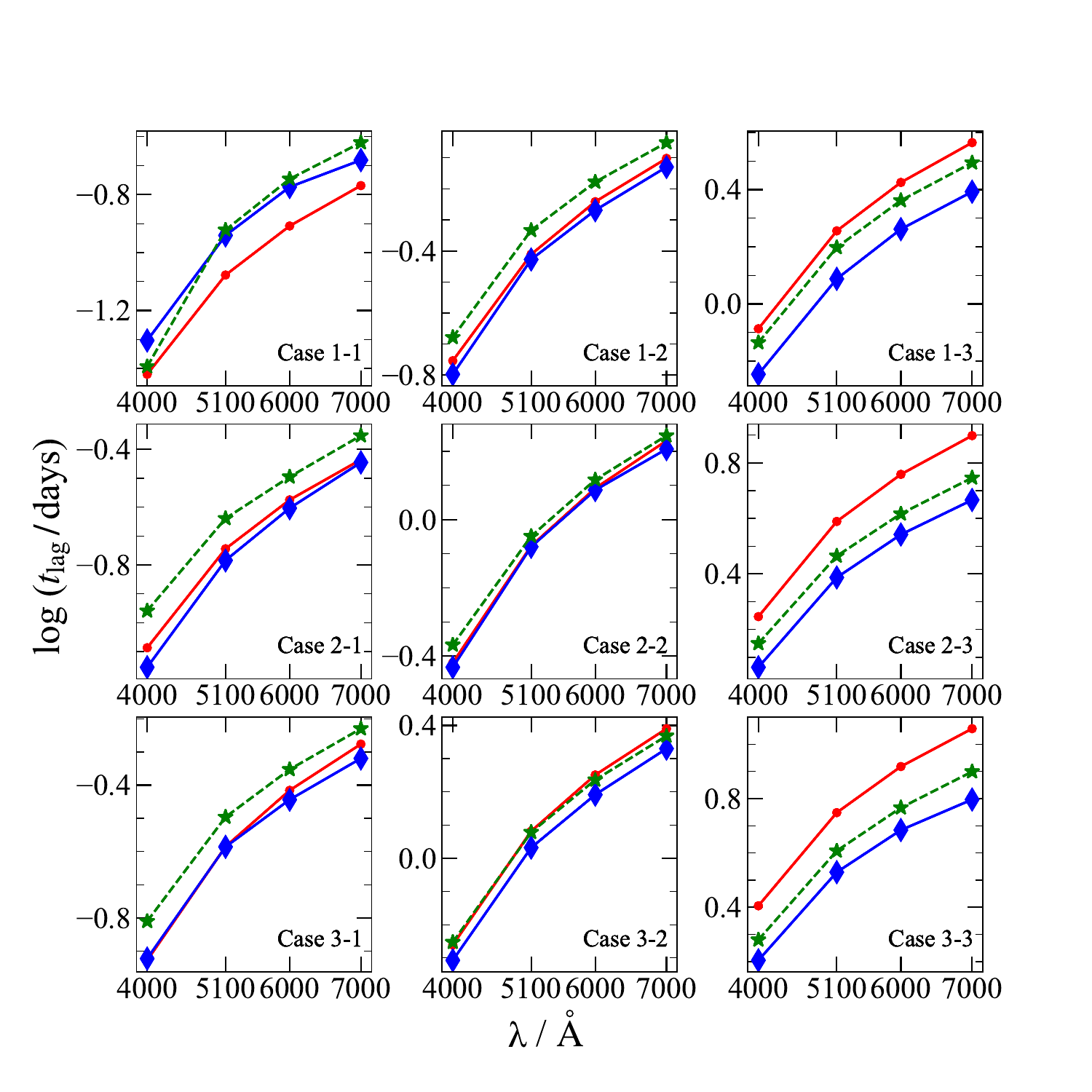}
\caption{Time lags as a function of wavelengths. The blue and red curves and symbols are taken from Figure \ref{fig3}. The green stars represent the inhomogeneous disk time lags for $f_{\mathrm{d}} = 0.1$ and $f_{\mathrm{c}} = 0.15$. For comparison, the results for the blue curves correspond to $f_{\mathrm{d}} = 0.3$ and $f_{\mathrm{c}} = 0.15$. The time lags decreases with increasing $f_{\mathrm{d}}$.}
\label{fig5}
\end{figure*}

\begin{figure*}[ht!]
\includegraphics[width=1 \textwidth]{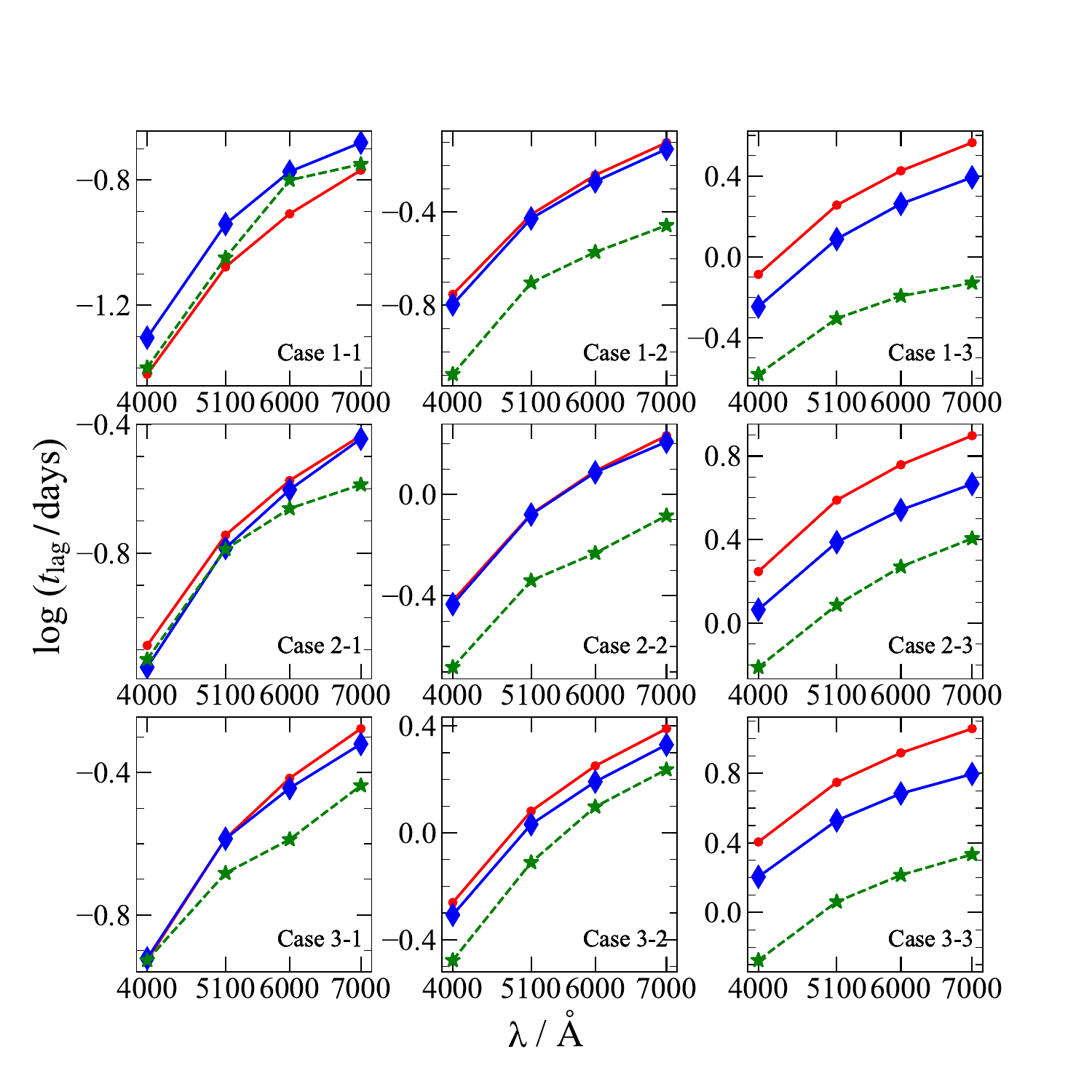}
\caption{Time lags as a function of wavelengths. The blue and red curves and symbols are taken from Figure \ref{fig3}. The green stars represent the inhomogeneous disk time lags for $f_{\mathrm{d}} = 0.3$ and $f_{\mathrm{c}} = 0.05$. For comparison, the results for the blue curves correspond to $f_{\mathrm{d}} = 0.3$ and $f_{\mathrm{c}} = 0.15$. The time lags decreases with decreasing $f_{\mathrm{c}}$.}
\label{fig6}
\end{figure*}

\subsection{Half-light radius } \label{sec:ratio} 

The inhomogeneous disk with or without X-ray illumination should have a half-light radius several times larger than that of an SSD. Figure \ref{fig2} shows the relationship between the size ratio of the half-light radius of the inhomogeneous disk with the X-ray illumination ($r_{\mathrm{h, r}}$) to the SSD model ($r_{\mathrm{h, S}}$) and the rest-frame wavelength for the nine cases in Table \ref{tab:Parameters}; the black dots and open circles are for $\tau_{\mathrm{DRW}}=1/(\alpha \Omega_{\mathrm{K}})$ (i.e., Eq.~\ref{eq:5}) and $\tau_{\mathrm{DRW}}=200$ days, respectively. For these 9 cases in Figure \ref{fig2}, the size ratio ranges from 2.5\,–\,4.2 \cite[also see][]{Dexter2011}, which is consistent with microlensing observations \citep[e.g.,][]{Morgan2010}. Comparing with $\tau_{\mathrm{DRW}}=200$ days, the size ratios with $\tau_{\mathrm{DRW}}=1/(\alpha \Omega_{\mathrm{K}})$ depend upon wavelength in a complex way. Unfortunately, the half-light radius differences between the two choices of $\tau_{\mathrm{DRW}}$ are too small to be distinguished by current microlensing observations.


\begin{figure*}[ht!]
\includegraphics[width=1 \textwidth]{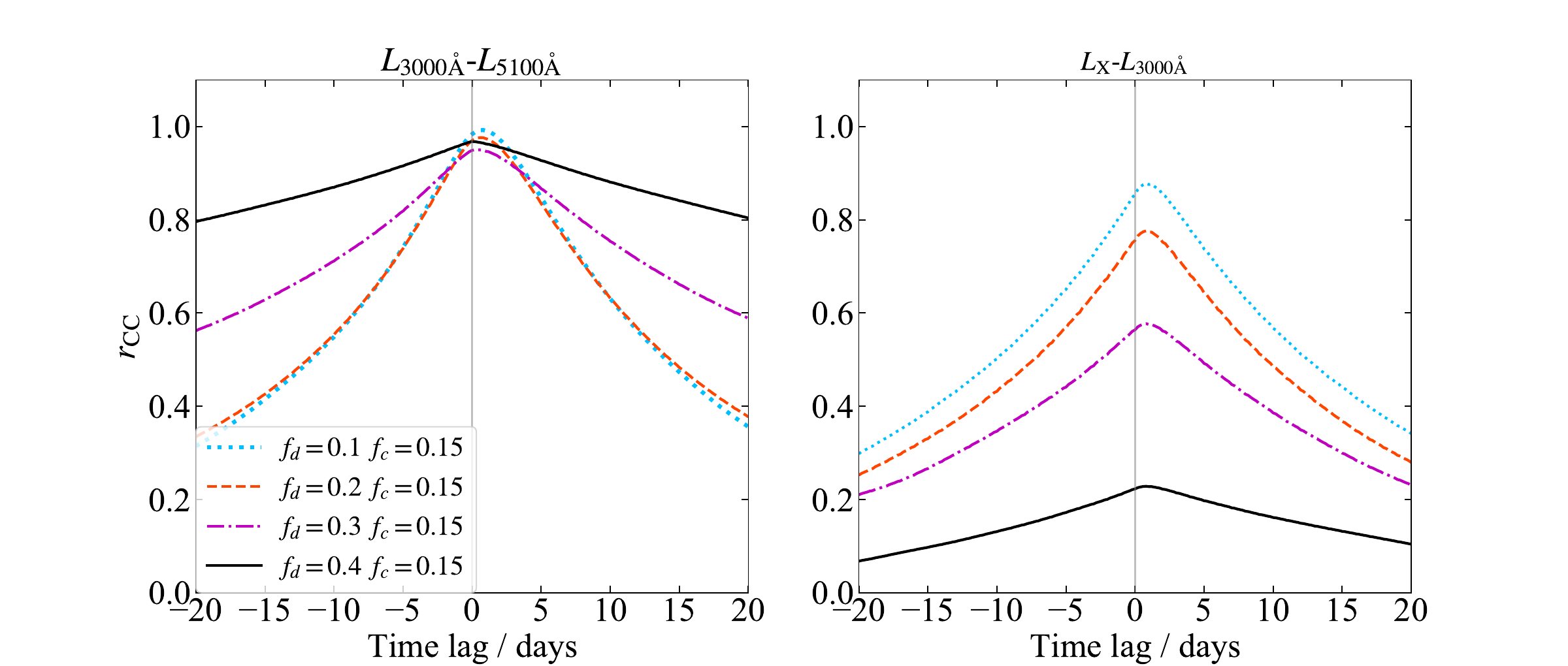}
\caption{The cross correlation functions. Left panel: the cross correlation functions between the $3000\ \mathrm{\AA}$ and $5100\ \mathrm{\AA}$ emission for the cases of $f_{\mathrm{d}}=0.1$, $0.2$, $0.3$, and $0.4$. In these cases, $f_{\mathrm{c}} = 0.15$. As the independent temperature-fluctuation amplitude ($f_{\mathrm{d}}$) increases, the peak and centroid of the function approaches zero time lag. Right panel: the CCF functions with time lags between light curves of X-ray and 3000$\textrm{\AA}$ emission. The peak and the correlation coefficient decrease with increasing $f_{\mathrm{d}}$. \label{fig7}}
\end{figure*}

\subsection{Time lag} \label{sec:tlag}

We calculate the time lags for the nine cases listed in Table~\ref{tab:Parameters}. We first consider the results for $\tau_{\mathrm{DRW}}=1/(\alpha \Omega_{\mathrm{K}})$. We use the ICCF method (Section~\ref{sec:time lag}) to obtain the time lags of the $4000\ \mathrm{\AA}$, $5100\ \mathrm{\AA}$, $6000\ \mathrm{\AA}$, and $7000\ \mathrm{\AA}$ emission with respect to the $3000\ \mathrm{\AA}$ emission. The time-lag searching ranges for the ICCF are from $-40$ days to $40$ days, with a uniform step of $0.2$ days. To match the real observational strategy, for each wavelength, we break the $25, 000$-day long light curve into $20$ non-overlapping segments, each segment lasts for $125$ days. Then, for each of the $4000\ \mathrm{\AA}$, $5100\ \mathrm{\AA}$, $6000\ \mathrm{\AA}$, and $7000\ \mathrm{\AA}$ emission, one can obtain $20$ time lags (with respect to the $3000\ \mathrm{\AA}$ emission) from the $20$ cross-correlation functions; the median values of the $20$ time lags are taken as our final results. 

Unlike the half-light radius result (see Section \ref{sec:ratio}), the interband time lags in the X-ray reprocessing of an inhomogeneous disk are not larger than those of an SSD. Figure~\ref{fig3} shows the median time lags of the X-ray reprocessing of the inhomogeneous disk (blue symbols) as a function of wavelength. For the sake of comparison, we also show the analytical time lags of the X-ray reprocessing of an SSD (red symbols; Eq.~\ref{eq:14}). For most cases, the interband time lags in the X-ray reprocessing of an inhomogeneous disk are even smaller than those of an SSD. Overall, the time-lag differences between the two models are rather small (less than 0.2 dex). Furthermore, the ratios of the two model lags decrease with increasing black hole mass or accretion rate, that is, the ratios and luminosity are inversely correlated. This is consistent with the observations of \cite{Li2021}. 

The time-lag results for other parameters are also considered. For instance, we now set $\tau_{\mathrm{DRW}}=200$ days (and leave other parameters unchanged) and calculate the interband time lags for the X-ray reprocessing of an inhomogeneous disk (Figure \ref{fig4}). We find that the time lags are significantly smaller than the results for $\tau_{\mathrm{DRW}}=1/(\alpha \Omega_{\mathrm{K}})$ (Figure \ref{fig3}). This is related to the fact that the damping timescale for $\tau_{\mathrm{DRW}}=1/(\alpha \Omega_{\mathrm{K}})$ are almost always significantly larger than $200$ days for the cases listed in Table~\ref{tab:Parameters}. Hence, the independent disk temperature fluctuations (i.e., the first term in the right hand of Eq.~\ref{eq:6}) for $\tau_{\mathrm{DRW}}=1/(\alpha \Omega_{\mathrm{K}})$ are significantly weaker than for $\tau_{\mathrm{DRW}}=200$ days. Then, we revise $f_{\mathrm{d}}$ to $0.1$ (leaving other parameters unchanged) and find that the time lags (Figure~\ref{fig5}) are slightly larger than the results for $f_{\mathrm{d}}=0.3$ (Figure~\ref{fig3}). Note that the independent disk temperature fluctuations decrease with decreasing $f_{\mathrm{d}}$. We also reduce $f_{\mathrm{c}}$ from $0.15$ to $0.05$ and find that the time lags also decrease. Hence, the interband time lags of the X-ray reprocessing of an inhomogeneous disk depend mostly upon the relative importance between the independent disk temperature fluctuations (i.e., the first term in the right hand of Eq.~\ref{eq:6}) and the X-ray induced coherent temperature fluctuations (i.e., the second term in the right hand of Eq.~\ref{eq:6}). 

To understand the dependence of the time lags upon the relative importance between the independent and X-ray induced coherent time fluctuations, we explore the CCFs for various of $f_{\mathrm{d}}$ (and fix other parameters). The CCFs between the $3000\ \mathrm{\AA}$ and $5100\ \mathrm{\AA}$ variations are shown in the left panel of Figure \ref{fig7} for $f_{\mathrm{d}}=0.1$, $0.2$, $0.3$, and $0.4$, respectively. It is evident that, as $f_{\mathrm{d}}$ increases, the CCF peak and centroid move towards the zero time lag. This is because of the following two reasons: first, if one increases $f_{\mathrm{d}}$, independent disk temperature fluctuations would contribute more to the flux variations; second, the CCF of the flux variations of two wavelengths driven by independent disk temperature fluctuations has a peak at the zero time lag \citep[see, e.g., Figure 18 in][]{Cai2016}.

\section{Discussion} \label{sec:discussion}


\subsection{The AGN disk size} \label{sec:discussion_size}

Similar to \cite{Dexter2011}, we confirm that the half-light radius of an inhomogeneous disk with the X-ray illumination is larger than that of the SSD by a factor of $\sim 3$. The half-light radii depend weakly upon the relationship between the damping timescale and the radius. Unfortunately, current microlensing measurements of quasar half-light radius have significant uncertainties and are therefore unable to distinguish the radius-dependent damping-timescale model from the constant-damping-timescale model. 

An interesting prediction of the X-ray processing of an inhomogeneous disk model is that, unlike the half-light radius, the interband time lags are roughly consistent with (or even smaller than) those of an SSD. The time lags depend upon the relative importance between the X-ray induced coherent and intrinsic independent disk temperature fluctuations. Hence, this model can explain recent disk reverberation mapping measurements of some quasars \citep[e.g.,][]{Kokubo2018, Homayouni2019, Yu2020} if independent disk temperature fluctuations do not dominate over X-ray induced coherent temperature fluctuations in these targets. The ratios between the interband time lags and the half-light radii, which can be simultaneously measured from the continuum reverberation mapping of gravitationally lensed quasars, provide an unique way to probe the intrinsic independent disk temperature fluctuations. 

MHD turbulence is widely accepted as a key process in the SMBH accretion since it can effectively remove the angular momentum of the accreted gas \citep[e.g.,][]{Balbus1991, Balbus2003}. Numerical simulations show evident temperature fluctuations in the accretion disk, which are ultimately driven by MHD turbulence \citep[e.g.,][]{Secunda2023}. Hence, the accretion disk itself is indeed inhomogeneous to some degree, and independent disk temperature fluctuations should contribute to the observed flux variations in AGNs. The time-lag predictions of the X-ray reprocessing of such a disk are incompatible with disk reverberation mapping of local AGNs \citep[e.g.,][]{Fausnaugh2016}. Alternative models \citep[e.g.,][]{Hall2018, Lawther2018, Sun2019, Kammoun2021, Zdziarski2022, Starkey2023} have been proposed to account for the larger-than-expected time lags in local AGNs. We stress that independent disk temperature fluctuations are expected to be large in these AGNs, since they have relatively low luminosities and small black-hole masses. Therefore, it is vital to account for the time-lag decrements caused by independent disk temperature fluctuations when comparing X-ray reprocessing models with disk reverberation mapping observations. 

\subsection{Cross correlations}

The cross-correlations between the X-ray and UV/optical emission can be weak in the X-ray reprocessing of an inhomogeneous disk. This is simply because the UV/optical variations in this model are caused by the extrinsic variable X-ray illumination and intrinsic independent disk temperature fluctuations. The correlation coefficients between the X-ray and UV/optical emission decrease with increasing the intrinsic, independent disk temperature fluctuations (the right panel Figure~\ref{fig7}). Hence, the X-ray reprocessing of an inhomogeneous disk provides a natural explanation for the weak correlations between X-ray and UV/optical variations observed in some AGNs \citep[e.g.,][]{Schimoia2015, Edelson2019}. \cite{Secunda2023} use 3D multi-frequency MHD simulations to simulate X-ray reverberation in the UV-emitting regions of AGN disks. The X-ray reprocessing drives small UV/optical variations on short timescales. Instead, the magnetorotational instability and convection generate significant intrinsic UV variability in their work. Their simulation shows the weak correlations between X-ray and UV-optical light curves, consistent with our result.

\subsection{How to increase continuum time lags}

There are several ways to increase the interband time lags in the X-ray reprocessing of an inhomogeneous disk model. For instance, the diffuse continuum from BLR clouds may contribute to the observed AGN continua and significantly enlarge the interband time lags \citep[e.g.,][]{Lawther2018}. Interestingly, there are indications that AGNs with larger optical variability amplitudes tend to show stronger broad emission lines \citep[e.g.,][]{Kang2021}. Alternatively, the disk size at a given wavelength can be increased because of strong winds \citep{Sun2019}, non-blackbody emission \citep{Hall2018}, a steep rim or rippled structures \citep{Starkey2023} or a large corona height \citep{Kammoun2021}. In these scenarios, the X-ray emission is assumed to be strong enough to drive UV/optical variability. 

The X-ray illumination might not be the only mechanism to drive interband time lags. Very recently, cross correlations with time lags have been detected in an X-ray weak quasar \citep{Marculewicz2023}, challenging the X-ray reprocessing scenario. \cite{Cai2018} suggest that a common large-scale variation can propagate across the disk at a speed slower than the speed of light and the response timescale of the temperature fluctuation to this common variation increases with radius. This model can nicely reproduce the UV/optical interband time lags and color variations in NGC 5548. Later, \cite{Cai2020} assume that the same common variation can also affect the X-ray corona and can account for the large time lags between X-ray emission and UV/optical emission, which are about tens times larger than the X-ray light travel timescales. The physical origin of the common temperature fluctuation is not yet known but may be disk thermal variations induced by UV emission \citep{Gardner2017}, slow axisymmetric temperature perturbations \citep{Neustadt2022}, or magnetic coupling \citep{Sun2020}. These possibilities will be explored in future studies.

\section{conclusions} \label{sec:conclusion}

We have presented the half-light radii and interband time lags of the X-ray reprocessing of an inhomogeneous AGN accretion disk. Our main results can be summarized as follows.
\begin{itemize}
    \item While the half-light radii are larger than those for SSD, the interband time lags are consistent or even smaller than SSD (Figures~\ref{fig3} to \ref{fig6}; Section~\ref{sec:tlag}). 
    \item The interband time lags decrease with increasing independent disk temperature fluctuations (Figure~\ref{fig7}; Section~\ref{sec:tlag}), and this effect should be properly considered in disk mapping modelings (Section~\ref{sec:discussion_size}). 
    \item The cross-correlation between the X-ray and disk emission can be very weak (Figure~\ref{fig7}), consistent with observations. 
\end{itemize}

The calculation presented in this work can be improved in several aspects. For instance, accretion-disk winds, which can enlarge the disk sizes \citep[e.g.,][]{Sun2019} and change the disk spectral energy distribution \citep[e.g.,][]{Slone2012}, should be taken into consideration. In addition, we also plan to consider the possible non-blackbody emission proposed by \cite{Hall2018}. These improvements will enable us to better test inhomogeneous accretion-disk models against quasar microlensing observations and continuum reverberation mappings.

\begin{acknowledgements}
We acknowledge support from the National Key R\&D Program of China (No. 2023YFA1607903). G.W.R. and M.Y.S. acknowledge support from the National Natural Science Foundation of China (NSFC-12322303), the Natural Science Foundation of Fujian Province of China (No. 2022J06002), and the China Manned Space Project grant (No. CMS-CSST-2021-A06). Z.Y.C. and J.X.W. acknowledge support from the National Natural Science Foundation of China (NSFC-12033006). G.W.R. and M.Y.S. thank Haikun Li for maintaining computing resources. We thank the referee for his/her useful comments that improve the manuscipt. 
\end{acknowledgements}




\begin{thebibliography}{}
\expandafter\ifx\csname natexlab\endcsname\relax\def\natexlab#1{#1}\fi
\providecommand{\url}[1]{\href{#1}{#1}}
\providecommand{\dodoi}[1]{doi:~\href{http://doi.org/#1}{\nolinkurl{#1}}}
\providecommand{\doeprint}[1]{\href{http://ascl.net/#1}{\nolinkurl{http://ascl.net/#1}}}
\providecommand{\doarXiv}[1]{\href{https://arxiv.org/abs/#1}{\nolinkurl{https://arxiv.org/abs/#1}}}

\bibitem[Balbus \& Hawley(1991)]{Balbus1991} Balbus, S.~A. \& Hawley, J.~F.\ 1991, \apj, 376, 214. \dodoi{10.1086/170270}

\bibitem[Balbus(2003)]{Balbus2003} Balbus, S.~A.\ 2003, \araa, 41, 555. \dodoi{10.1146/annurev.astro.41.081401.155207}

\bibitem[Blackburne et al.(2011)]{Blackburne2011} Blackburne, J.~A., Pooley, D., Rappaport, S., et al.\ 2011, \apj, 729, 34. \dodoi{10.1088/0004-637X/729/1/34}

\bibitem[Cackett et al.(2007)]{Cackett2007} Cackett, E.~M., Horne, K., \& Winkler, H.\ 2007, \mnras, 380, 669. \dodoi{10.1111/j.1365-2966.2007.12098.x}

\bibitem[Cackett et al.(2022)]{Cackett2022} Cackett, E.~M., Zoghbi, A., \& Ulrich, O.\ 2022, \apj, 925, 29. \dodoi{10.3847/1538-4357/ac3913}

\bibitem[Cai et al.(2016)]{Cai2016} Cai, Z.-Y., Wang, J.-X., Gu, W.-M., et al.\ 2016, \apj, 826, 7. \dodoi{10.3847/0004-637X/826/1/7}

\bibitem[Cai et al.(2018)]{Cai2018} Cai, Z.-Y., Wang, J.-X., Zhu, F.-F., et al.\ 2018, \apj, 855, 117. \dodoi{10.3847/1538-4357/aab091}

\bibitem[Cai et al.(2020)]{Cai2020} Cai, Z.-Y., Wang, J.-X., \& Sun, M.\ 2020, \apj, 892, 63. \dodoi{10.3847/1538-4357/ab7991}

\bibitem[Cai \& Wang(2023)]{Cai2023} Cai, Z.-Y. \& Wang, J.-X.\ 2023, Nature Astronomy. \dodoi{10.1038/s41550-023-02088-5}

\bibitem[Capellupo et al.(2015)]{Capellupo2015} Capellupo, D.~M., Netzer, H., Lira, P., et al.\ 2015, \mnras, 446, 3427. \dodoi{10.1093/mnras/stu2266}

\bibitem[Dai et al.(2010)]{Dai2010} Dai, X., Kochanek, C.~S., Chartas, G., et al.\ 2010, \apj, 709, 278. \dodoi{10.1088/0004-637X/709/1/278}

\bibitem[Dexter \& Agol(2011)]{Dexter2011} Dexter, J. \& Agol, E.\ 2011, \apjl, 727, L24, \dodoi{10.1088/2041-8205/727/1/L24}

\bibitem[Edelson et al.(2019)]{Edelson2019} Edelson, R., Gelbord, J., Cackett, E., et al.\ 2019, \apj, 870, 123. \dodoi{10.3847/1538-4357/aaf3b4}

\bibitem[Fausnaugh et al.(2016)]{Fausnaugh2016} Fausnaugh, M.~M., Denney, K.~D., Barth, A.~J., et al.\ 2016, \apj, 821, 56. \dodoi{10.3847/0004-637X/821/1/56}

\bibitem[Fausnaugh et al.(2018)]{Fausnaugh2018} Fausnaugh, M.~M., Starkey, D.~A., Horne, K., et al.\ 2018, \apj, 854, 107. \dodoi{10.3847/1538-4357/aaaa2b}

\bibitem[Fian et al.(2023)]{Fian2023} Fian, C., Chelouche, D., \& Kaspi, S.\ 2023, arXiv:2307.14824. \dodoi{10.48550/arXiv.2307.14824}

\bibitem[Gardner \& Done(2017)]{Gardner2017} Gardner, E. \& Done, C.\ 2017, \mnras, 470, 3591. \dodoi{10.1093/mnras/stx946}

\bibitem[Gaskell \& Sparke(1986)]{Gaskell1986} Gaskell, C.~M. \& Sparke, L.~S.\ 1986, \apj, 305, 175. \dodoi{10.1086/164238}

\bibitem[Hall et al.(2018)]{Hall2018} Hall, P.~B., Sarrouh, G.~T., \& Horne, K.\ 2018, \apj, 854, 93. \dodoi{10.3847/1538-4357/aaa768}

\bibitem[Homayouni et al.(2019)]{Homayouni2019} Homayouni, Y., Trump, J.~R., Grier, C.~J., et al.\ 2019, \apj, 880, 126. \dodoi{10.3847/1538-4357/ab2638}

\bibitem[Hook et al.(1994)]{Hook1994} Hook, I.~M., McMahon, R.~G., Boyle, B.~J., et al.\ 1994, \mnras, 268, 305. \dodoi{10.1093/mnras/268.2.305}

\bibitem[Jiang et al.(2017)]{Jiang2017} Jiang, Y.-F., Green, P.~J., Greene, J.~E., et al.\ 2017, \apj, 836, 186. \dodoi{10.3847/1538-4357/aa5b91}

\bibitem[Kammoun et al.(2021)]{Kammoun2021} Kammoun, E.~S., Papadakis, I.~E., \& Dov{\v{c}}iak, M.\ 2021, \mnras, 503, 4163. \dodoi{10.1093/mnras/stab725}

\bibitem[Kang et al.(2021)]{Kang2021} Kang, W.-Y., Wang, J.-X., Cai, Z.-Y., et al.\ 2021, \apj, 911, 148. \dodoi{10.3847/1538-4357/abeb69}

\bibitem[Kokubo(2018)]{Kokubo2018} Kokubo, M.\ 2018, \pasj, 70, 97. \dodoi{10.1093/pasj/psy096}

\bibitem[Koz{\l}owski(2017)]{Kozowski2017} Koz{\l}owski, S.\ 2017, \aap, 597, A128. \dodoi{10.1051/0004-6361/201629890}

\bibitem[Lawther et al.(2018)]{Lawther2018} Lawther, D., Goad, M.~R., Korista, K.~T., et al.\ 2018, \mnras, 481, 533. \dodoi{10.1093/mnras/sty2242}

\bibitem[Li et al.(2021)]{Li2021} Li, T., Sun, M., Xu, X., et al.\ 2021, \apjl, 912, L29. \dodoi{10.3847/2041-8213/abf9aa}

\bibitem[Li et al.(2019)]{Li2019} Li, Y.-P., Yuan, F., \& Dai, X.\ 2019, \mnras, 483, 2275. \dodoi{10.1093/mnras/sty3245}

\bibitem[Marculewicz et al.(2023)]{Marculewicz2023} Marculewicz, M., Sun, M., Wu, J., et al.\ 2023, \apj, 956, 126. \dodoi{10.3847/1538-4357/acf312}

\bibitem[MacLeod et al.(2010)]{MacLeod2010} MacLeod, C.~L., Ivezi{\'c}, {\v{Z}}., Kochanek, C.~S., et al.\ 2010, \apj, 721, 1014. \dodoi{10.1088/0004-637X/721/2/1014}

\bibitem[Maoz et al.(2002)]{Maoz2002} Maoz, D., Markowitz, A., Edelson, R., et al.\ 2002, \aj, 124, 1988. \dodoi{10.1086/342937}

\bibitem[Morgan et al.(2010)]{Morgan2010} Morgan, C.~W., Kochanek, C.~S., Morgan, N.~D., et al.\ 2010,\apj, 712, 1129, \dodoi{10.1088/0004-637X/712/2/1129}

\bibitem[Morgan et al.(2018)]{Morgan2018} Morgan, C.~W., Hyer, G.~E., Bonvin, V., et al.\ 2018, \apj, 869, 106. \dodoi{10.3847/1538-4357/aaed3e}

\bibitem[Nayakshin \& Kazanas(2002)]{Nayakshin2002} Nayakshin, S. \& Kazanas, D.\ 2002, \apj, 567, 85. \dodoi{10.1086/338333}

\bibitem[Neustadt \& Kochanek(2022)]{Neustadt2022} Neustadt, J.~M.~M. \& Kochanek, C.~S.\ 2022, \mnras, 513, 1046. \dodoi{10.1093/mnras/stac888}

\bibitem[Novikov \& Thorne(1973)]{Novikov1973} Novikov, I.~D. \& Thorne, K.~S.\ 1973, Black Holes (Les Astres Occlus), 343

\bibitem[Panagiotou et al.(2022)]{Panagiotou2022} Panagiotou, C., Kara, E., \& Dov{\v{c}}iak, M.\ 2022, \apj, 941, 57. \dodoi{10.3847/1538-4357/aca2a4}

\bibitem[Peterson et al.(1998)]{Peterson1998} Peterson, B.~M., Wanders, I., Horne, K., et al.\ 1998, \pasp, 110, 660. \dodoi{10.1086/316177}

\bibitem[Peterson et al.(2004)]{Peterson2004} Peterson, B.~M., Ferrarese, L., Gilbert, K.~M., et al.\ 2004, \apj, 613, 682. \dodoi{10.1086/423269}

\bibitem[Pooley et al.(2007)]{Pooley2007} Pooley, D., Blackburne, J.~A., Rappaport, S., et al.\ 2007, \apj, 661, 19, \dodoi{10.1086/512115}

\bibitem[Schimoia et al.(2015)]{Schimoia2015} Schimoia, J.~S., Storchi-Bergmann, T., Grupe, D., et al.\ 2015, \apj, 800, 63. \dodoi{10.1088/0004-637X/800/1/63}

\bibitem[Secunda et al.(2023)]{Secunda2023} Secunda, A., Jiang, Y.-F., \& Greene, J.~E.\ 2023, arXiv:2311.10820. \dodoi{10.48550/arXiv.2311.10820}

\bibitem[Shakura \& Sunyaev(1973)]{Shakura1973} Shakura, N.~I. \& Sunyaev, R.~A.\ 1973, \aap, 500, 33

\bibitem[Slone \& Netzer(2012)]{Slone2012} Slone, O. \& Netzer, H.\ 2012, \mnras, 426, 656. \dodoi{10.1111/j.1365-2966.2012.21699.x}

\bibitem[Starkey et al.(2023)]{Starkey2023} Starkey, D.~A., Huang, J., Horne, K., et al.\ 2023, \mnras, 519, 2754. \dodoi{10.1093/mnras/stac3579}

\bibitem[Stone et al.(2022)]{Stone2022} Stone, Z., Shen, Y., Burke, C.~J., et al.\ 2022, \mnras, 514, 164. \dodoi{10.1093/mnras/stac1259}

\bibitem[Suberlak et al.(2021)]{Suberlak2021} Suberlak, K.~L., Ivezi{\'c}, {\v{Z}}., \& MacLeod, C.\ 2021, \apj, 907, 96. \dodoi{10.3847/1538-4357/abc698}

\bibitem[Sun et al.(2014)]{Sun2014} Sun, Y.-H., Wang, J.-X., Chen, X.-Y., et al.\ 2014, \apj, 792, 54. \dodoi{10.1088/0004-637X/792/1/54}

\bibitem[Sun et al.(2018)]{Sun2018} Sun, M., Grier, C.~J., \& Peterson, B.~M.\ 2018, Astrophysics Source Code Library. ascl:1805.032

\bibitem[Sun et al.(2019)]{Sun2019} Sun, M., Xue, Y., Trump, J.~R., et al.\ 2019, \mnras, 482, 2788. \dodoi{10.1093/mnras/sty2885}

\bibitem[Sun et al.(2020)]{Sun2020} Sun, M., Xue, Y., Guo, H., et al.\ 2020, \apj, 902, 7. \dodoi{10.3847/1538-4357/abb1c4}

\bibitem[Tie \& Kochanek(2018)]{Tie2018} Tie, S.~S. \& Kochanek, C.~S.\ 2018, \mnras, 473, 80. \dodoi{10.1093/mnras/stx2348}

\bibitem[Ulrich et al.(1997)]{Ulrich1997} Ulrich, M.-H., Maraschi, L., \& Urry, C.~M.\ 1997, \araa, 35, 445. \dodoi{10.1146/annurev.astro.35.1.445}

\bibitem[Yu et al.(2020)]{Yu2020} Yu, Z., Martini, P., Davis, T.~M., et al.\ 2020, \apjs, 246, 16. \dodoi{10.3847/1538-4365/ab5e7a}

\bibitem[Zdziarski et al.(2022)]{Zdziarski2022} Zdziarski, A.~A., You, B., \& Szanecki, M.\ 2022, \apjl, 939, L2. \dodoi{10.3847/2041-8213/ac9474}

\bibitem[Zhou et al.(2024)]{Zhou2024} Zhou, S., Sun, M., Cai, Z.-Y., et al.\ 2024, arXiv:2403.01691. \dodoi{10.48550/arXiv.2403.01691}

\bibitem[Zu et al.(2013)]{Zu2013} Zu, Y., Kochanek, C.~S., Koz{\l}owski, S., et al.\ 2013, \apj, 765, 106. \dodoi{10.1088/0004-637X/765/2/106}




\end{thebibliography}
\end{document}